\documentclass[twocolumn,showpacs,preprintnumbers,amsmath,amssymb]{revtex4}

\usepackage{graphicx}% Include figure files

\usepackage{dcolumn}% Align table columns on decimal point

\usepackage{bm}% bold math

\begin{document}

\preprint{APS/123-QED}

\title{Electric fields at the quark surface of strange stars
in the color-flavor locked phase}

\author{Vladimir V. Usov}

\affiliation{Department of Condensed Matter Physics, Weizmann
Institute of Science, Rehovot 76100, Israel \\}

\date{\today}% It is always \today, today,

             %  but any date may be explicitly specified

\begin{abstract}

It is shown that extremely strong electric fields may be generated
at the surface of strange quark matter in the color-flavor locked
phase because of the surface effects. Some properties of strange
stars made of this matter are briefly discussed.

\end{abstract}

% PACS, the Physics and Astronomy

                             % Classification Scheme.

%\keywords{Suggested keywords}%Use showkeys class option if keyword

                              %display desired

\maketitle

\section{\label{sec:level1}INTRODUCTION}

Strange quark matter (SQM) that consists of deconfined $u$, $d$,
and $s$ quarks may be the absolute ground state of the strong
interaction, i.e., absolutely stable with respect to $^{56}$Fe
\cite{W84,FJ84}. If SQM is approximated as noninteracting quarks,
chemical equilibrium with respect to the weak interaction together
with the relatively large mass of the $s$ quark imply that the $s$
quarks are less abundant than the other quarks, and electrons are
required in SQM to neutralize the electric charge of the quarks.

The electron density at vanishing pressure is $\sim 10^{-4}$ of
the quark density \cite{AFO86,KWWG95}. The electrons, being bound
to SQM by the electromagnetic interaction alone, are able to move
freely across the SQM surface, but clearly cannot move to infinity
because of the bulk electrostatic attraction to the quarks. The
distribution of electrons extends several hundred fermis above the
SQM surface, and an enormous electric field $E\simeq 5\times
10^{17}$ V~cm$^{-1}$ is generated in the surface electron layer to
prevent the electrons from escaping to infinity, counterbalancing
the degeneracy and thermal pressure \cite{AFO86,KWWG95,HX02}.

Strange stars made entirely of SQM have long been proposed as an
alternative to neutron stars (e.g., \cite{W84,AFO86}). The
electric field at the surface of strange stars drastically affects
their observational appearance. The point is that this field is a
few 10 times higher than the critical field $E_{\rm cr}\simeq
1.3\times 10^{16}$ V~cm$^{-1}$ at which vacuum is unstable to
creation of $e^+e^-$ pairs. Therefore, a hot strange star with a
bare SQM surface may be a powerful source of $e^+e^-$ pairs which
are created in the surface electric field and flow away from the
star \cite{U98,U01,AMU03}. The thermal luminosity of such a star
in pairs depends on the surface temperature $T_S$ and may be as
high as $\sim 10^{51}$ ergs~s$^{-1}$ at the moment of formation
when $T_{\rm S}$ may be up to $\sim 10^{11}$~K. The luminosity of
a young bare strange star in pairs may remain high enough
($\gtrsim 10^{36}$ ergs~s$^{-1}$) for $\sim 10^3$ yr \cite{PU02}.

The surface electric field may be also responsible for existence
of a crust of "normal" matter (ions and electrons) at the SQM
surface of a strange star \cite{AFO86}. The ions in the inner
layer are supported against the gravitational attraction to the
underlying strange star by the electric field.

It is becoming widely accepted that because of an attractive
interaction between quarks in some specific channels, the ground
state of SQM is a color superconductor \cite{ABR01}. At asymptotic
densities ($\gg n_0$), this superconductor is likely to be in the
color-flavor locked (CFL) phase in which quarks of all three
flavors and three colors are paired in a single condensate, where
$n_0\simeq 0.16$ fm$^{-3}$ is the normal nuclear density.
Unfortunately, at intermediate densities ($\sim 2 n_0$) that are
relevant to the surface layers of strange stars, the QCD phase of
SQM is uncertain. In this low density regime, the SQM may be not
only in the CFL phase, but also in the "two color-flavor
superconductor" (2SC) phase in which only $u$ and $d$ quarks of
two color are paired in a single condensate, while the ones of
third color and the s quarks of all three colors are unpaired.

However, it was recently argued that the density range where the
2SC phase may exist is small, if any \cite{AR02}. In the CFL and
2SC phases, the Cooper pairs are made of quarks with equal and
opposite momenta. Another possibility is a crystalline color
superconductor (CCS), which involves pairing between quarks whose
momenta do not add to zero \cite{BR02}. In the 2SC and CCS phases,
electrons are present, and the electron density is more or less
the same as in the unpaired SQM. Therefore, superconducting
strange stars with the SQM surface layers in one of these phases
are expected to be rather similar to strange stars that are made
of noninteracting, unpaired quarks. It is now commonly accepted
that the CFL phase of SQM in bulk consists of equal numbers of
$u$, $d$, and $s$ quarks and is electrically neutral in the
absence of any electrons \cite{RW01}. At first sight, an extremely
strong electric field is absent at the surface of SQM in the CFL
phase. In turn, this could result in qualitatively changes of many
properties of strange stars (such as absence of both the intence
emission of $e^+e^-$ pairs and normal matter crusts) if the
strange star surface is in the CFL phase. In this Letter, we show
that this is not the fact, and supercritical electric fields may
be generated at the CFL-phase surface because of the surface
effects.

\section{\label{sec:level2}ELECTRIC FIELDS AT THE CFL SURFACE}

The modification of the density of quark states near the boundary
of SQM gives rise to rather large positive surface tension for
massive quarks \cite{BJ87}. Other sources of surface tension are
probably smaller \cite{FJ84}, and we ignore them here.

The reason for the electrical neutrality of the CFL phase in bulk
is that BCS-like pairing minimizes the energy if the quark Fermi
momenta are equal. In turn, for equal Fermi momenta, the numbers
of $u$, $d$, $s$ quarks are equal, and the electric charge of the
quarks is zero. The properties of the CFL phase with taking into
account the surface effects were discussed by Madsen \cite{M01},
and it was shown that the number of massive quarks is reduced near
the boundary relative to the number of massless quarks at fixed
Fermi momenta. The change in number of quarks of flavor $i$ per
unit area is \cite{M01}
\begin{equation}
n_{i,S}= - {3\over 4\pi}p_{F,i}^2 \left[{1\over 2}+{\lambda_i\over
\pi}- {1\over\pi}(1+\lambda^2_i) \tan^{-1}(\lambda_i^{-1})\right],
\label{niS}
\end{equation}
where $i=\{u,d,s\}$, $p_{F,i}$ is the Fermi momentum of quarks of
flavor $i$, $\lambda_i=m_i/p_{F,i}$, and $m_i$ is the rest mass of
quarks of flavor $i$. The value of $n_{i,S}$ is always negative,
approaching zero for $\lambda_i\rightarrow 0$ (massless quarks).
The rest masses of $u$ and $d$ quarks are very small, and their
densities are not modified significantly by the surface. Thus, the
only appreciable contribution to the surface corrections arises
from the $s$ quarks, i.e., surface effects are highly flavor
dependent. Because of surface depletion of $s$ quarks thin layers
at the surface of the CFL phase are no longer electrically neutral
as in bulk. The charge per unit area is positive and equals
\begin{equation}
\sigma = -{1\over 3}en_{s,S}\,. \label{sigma}
\end{equation}

The thickness of the charged layer at the surface of SQM in the
CFL phase is of order of 1~fm, which is a typical strong
interaction length scale.

Electrons are required to neutralize electrically the charged
layer of SQM. The thickness of the electron distribution is about
two order more than the thickness of the SQM charged layer (see
\cite{AFO86,KWWG95} and below), and therefore, we assume that the
last is infinitesimal. In this case, the density of electrons
$n_e$ and the electrostatic potential $V$ are symmetric to the SQM
charged layer, i.e, $n_e (-z)=n_e(z)$ and $V(-z)=V(z)$, where $z$
is a space coordinate measuring height above the SQM surface. In
turn, the electric field $E=-dV/dz$ is directed from the SQM
charged layer, and $E(-z)=-E(z)$. The strength of the electric
field at the SQM surface ($|z| \rightarrow 0$) is
$E_0=[E(+0)-E(-0)]/2=2\pi\sigma$, where $\sigma $ is given by
equation (\ref{sigma}). Taking $m_s\simeq 150$ MeV and $p_{F,s}
\simeq 300$ MeV as typical parameters of SQM, from equations
(\ref{niS}) and (\ref{sigma}) we have $E_0\simeq 2.7\times
10^{18}\,{\rm V~cm}^{-1} \simeq 200E_{\rm cr}$ that is $\sim 5$
times larger than the surface electric field calculated for SQM in
the unpaired phase neglecting the surface effects (e.g.,
\cite{AFO86,KWWG95,HX02}). We hope to deal with these effects for
the unpaired, 2SC and CCS phases elsewhere.

In a simple Thomas-Fermi model, the distribution of the
electrostatic potential $V(z,T_{\rm S})$ near the SQM surface with
the temperature $T_S$ is described by Poisson's equation (e.g.,
\cite{AFO86,KWWG95,CH03})
\begin{equation}
{d^2V\over dz^2}={4\alpha\over 3\pi}(e^2V^3 +\pi^2T_S^2V)\,,
\label{d2V}
\end{equation}
where $\alpha$ is the fine-structure constant. The boundary
conditions for equation (\ref{d2V}) are
\begin{equation}
dV/dz =\mp\, 2\pi \sigma \label{d1V}
\end{equation}
at the external ($z=+ 0$) and internal ($z=- 0$) sides of the SQM
surface, respectively, and $V\rightarrow 0$ as $z\rightarrow \pm
\infty$. The first integral of equation (\ref{d2V}), which
satisfies the boundary condition at $z\rightarrow \pm\infty$, is
\begin{equation}
{dV\over dz}=\mp\left({2\alpha\over 3\pi}\right)^{1/2}
(e^2V^4+2\pi^2T^2_SV^2)^{1/2}\,. \label{d1Vint}
\end{equation}
where the sign $-$ or $+$ has to be taken at $z>0$ or $z<0$,
respectively. Using the boundary condition (\ref{d1V}) and
equations (\ref{sigma}) and (\ref{d1Vint}), we have the
electrostatic potential at the SQM surface
\begin{equation}
V(0, T_{\rm S})=V(0,0) \{[1+({T_S/ T_*})^4]^{1/2}- ({T_S/
T_*})^2]\}^{1/2},
\end{equation}
where
\begin{equation}
V(0,0)=\left({2\pi^3n_{s,S}^2 \over 3\alpha}\right)^{1/4}
\,\,\,{\rm and}\,\,\,\,\, T_*={eV(0,0)\over \pi}.
\end{equation}
or numerically
\begin{equation}
V(0,0)\simeq 3.6\times 10^7~{\rm V}\,\,\,\,\,\,{\rm and}\,\,\,\,\,
\,T_*\simeq 11.6~{\rm MeV}
\end{equation}
for $m_s\simeq 150$ MeV and $p_{F,s}\simeq 300$ MeV.

In the case of rather low temperatures ($T_S\ll T_*$), from
equation (\ref{d1Vint}) the electrostatic potential is
\begin{equation}
V(z,0)= \left({3\pi\over 2\alpha}\right)^{1/2} {1\over
e(|z|+z_0)}\,,
\end{equation}
where $z_0=(3\pi/2\alpha)^{1/2}[eV(0,0)]^{-1}\simeq 2\times 10^2$
fm is the typical thickness of the surface electron layer with a
strong electric field. In this layer, the number density of
electrons is (e.g., \cite{KWWG95,CH03})
\begin{equation}
n_e(z,0)={e^3V^3(z,0)\over 3\pi^2}={1\over 3\pi^2}
\left({3\pi\over 2\alpha}\right)^{3/2} {1\over (|z|+z_0)^3}\,.
\end{equation}

Hence, at the surface of SQM in the CFL phase a Coulomb barrier
with an extremely strong electric field may be present because of
the surface effects in spite of that this phase in bulk consists
of equal numbers of $u$, $d$, and $s$ quarks and is electrically
neutral in the absence of any electrons. In this case the electric
field is directed from the SQM surface and $E(-z)=-E(z)$. It is
worth noting that the pairing energy contribution has been
neglected by Madsen \cite{M01} in the derivation of equation (1).
However, since the pairing energy is rather small compared with
the other contributions to the energy, we think that this
approximation doesn't affect the conclusions in this paper at
least when the pairing energy is not extremely high.

\section{\label{sec:level3}ASTROPHYSICAL CONSEQUENCES}

A strange star at the moment of its formation may be very hot, and
the rate of neutrino-induced mass ejection from the stellar
surface may be very high \cite{WB92}. In this case, in a few
second after the star formation the normal-matter crust is blown
away, and the SQM surface is nearly (or completely) bare
\cite{U01}. We have shown that a Coulomb barrier with the
electrostatic potential of $\sim 3.6\times 10^7$~V may be present
at the bare SQM surface of a strange star in the CFL phase. Such a
barrier may be a powerful source of $e^+e^-$ pairs created in its
extremely strong electric field \cite{U98}. The strange star
luminosity in pairs remains high enough ($\gtrsim 10^{36}$
ergs~s$^{-1}$) as long as the surface temperature is higher than
$\sim 6\times 10^8$~K \cite{U01}. Below this temperature, $T_S<
6\times 10^8$~K, non-equilibrium quark-quark \cite{CH03} and
electron-electron \cite{JGPP04} bremsstrahlung radiation dominates
in the thermal emission from the surface of a bare strange star,
i.e., the pair production by the Coulomb barrier is not
significant.

At the surface of a strange star in the CFL phase a massive normal
matter crust may form by accretion of matter onto the star
\cite{AFO86,U97}. For the Coulomb barrier at the surface of such a
star the electrostatic potential of electrons is $eV(0,0) \simeq
36$ MeV that is more than the electron chemical potential ($\sim
25$ MeV) at which neutron drip occurs \cite{BPS71}. Therefore, the
maximum density of the crust is limited by neutron drip and is
about $4.3\times 10^{11}$ g~cm$^{-3}$ \cite{AFO86}. At this
density free neutrons begin to drip out of the most stable
nucleus, $^{118}$Kr ($Z=36$). Being electrically charge neutral,
the neutrons can gravitate toward the star's quark core where they
are converted into SQM. For a strange star in the CFL phase the
maximum mass of the crust is $\sim 10^{-5}M_\odot$. A massive
($\Delta M\sim 10^{-5}M_\odot$) crust of normal matter completely
obscures the star's SQM core, and in the observed mass range
($1<M/M_\odot <2$) it is difficult to discriminate between neutron
stars and strange stars with such crusts \cite{HZS86,G97,PPLS00}
(however, cf. \cite{P91,LB99}).

\begin{acknowledgments}

I am grateful to K. Rajagopal for valuable discussions. I thank
the referees for useful comments. This work was supported by the
Israel Science Foundation of the Israel Academy of Sciences and
Humanities.

\end{acknowledgments}

\newpage

\bibliography{apssamp}% Produces the bibliography via BibTeX.

%%%%%%%%%%%%%%%%%%%%%%%%%%%%%%%%%%%%%%%%%%%%%%%%%%%%%%%%%%%%%%%%%%%%%

%%%%%%%%%%%%%%%%%%%%%%%%%%%%%%%%%%%%%%%%%%%%%%%%%%%%%%%%%%%%%%%%%%%%%

%%%%%%%%%%%%%%%%%%%%%%%%%%%%%%%%%%%%%%%%%%%%%%%%%%%%%%%%%%%%%%%%%%%%

\end{document}